\documentclass[preprint,superscriptaddress,showpacs,preprintnumbers,amsmath,amssymb,aps,prl]{revtex4}
\usepackage{mathptmx}
\usepackage{graphicx}
\usepackage{color}
\begin{document}
\title{Order parameter fluctuations in natural time and $b$-value variation before large earthquakes}

\author{P. A. Varotsos}\thanks{{\bf Correspondence to:} P. Varotsos (pvaro@otenet.gr)}
\affiliation{Solid State Section and Solid Earth Physics
Institute, Physics Department, University of Athens,
Panepistimiopolis, Zografos 157 84, Athens, Greece}
\author{N. V. Sarlis}
\affiliation{Solid State Section and Solid Earth Physics
Institute, Physics Department, University of Athens,
Panepistimiopolis, Zografos 157 84, Athens, Greece}
\author{E. S. Skordas}
\affiliation{Solid State Section and Solid Earth Physics
Institute, Physics Department, University of Athens,
Panepistimiopolis, Zografos 157 84, Athens, Greece}

\newcommand {\equald} {\ {\raise-.5ex\hbox{$\buildrel d \over =$}}\ }

\begin{abstract}
Self-similarity may stem from two origins: the process' increments
infinite variance and/or process' memory. The $b$-value of the Gutenberg-Richter law
comes from the first origin. In the frame of natural
time analysis of earthquake data, a fall of the $b$-value observed
before large earthquakes reflects an increase of the order parameter
fluctuations upon approaching the critical point (mainshock).
The increase of these fluctuations, however, is also influenced
from the second origin of self-similarity, i.e., temporal
correlations between earthquake magnitudes. This is supported by
observations and simulations of an earthquake model.
\end{abstract}


 \maketitle

A large variety of natural systems exhibit irregular and complex
behavior which at first look seems to be erratic, but in fact
possesses scale-invariant structure, for example see Refs.
\cite{PEN95,KAL05}. A stochastic process $X(t)$ is called
self-similar\cite{LAM62} with index $H>0$ if it has the property
\begin{equation}
\label{selfsim} X(\lambda t) \equald \lambda^H X(t)  \quad\quad
\forall\quad  \lambda > 0.
\end{equation}
where the equality concerns the finite-dimensional distributions
of the process $X(t)$ on the right- and the left-hand side of the
equation ({\em not} the values of the process).

A point of crucial importance in analyzing data from complex
systems that exhibit scale-invariant structure, is the following:
In several systems this nontrivial structure stems from long-range
{\em temporal} correlations; in other words, the self-similarity
originates from the process' memory {\em only}. This is the case
for example of fractional Brownian motion. Alternatively, the
self-similarity may solely come from the process' increments {\em
infinite} variance. Such an example is L\'{e}vy stable motion (the
variance of L\'{e}vy stable distributions is infinite since they
have heavy tails\cite{WER05}, thus differing greatly from the
Gaussian ones). In general, however, the self-similarity may
result from both these origins\cite{KAN02}, the presence of which
can be in principle identified when analyzing the complex time
series in terms of the new time domain termed natural
time\cite{SPRINGER}.

The evolution of seismicity is a typical example of complex time
series. Several traditional studies were focused on the variation
of the $b$-value of the Gutenberg-Richter (G-R) law\cite{GUT54},
which states that the (cumulative) number of earthquakes with
magnitude greater than (or equal to) M, $N(\geq {\rm M})$,
occurring in a specified area and time is given by
\begin{equation}
 N(\geq {\rm M})=10^{a-b {\rm M}}, \label{GRorig}
 \end{equation}
where $b$ is a constant, varying only slightly from region to
region and the constant $a$ gives the logarithm of the number of
earthquakes with magnitude greater than zero\cite{SHC04}. These
studies found that the $b$-value decreases before a large event,
e.g., see Ref.\cite{LI78} (cases where $b$-value increases prior
to and then decreases sharply before a large event have been also
reported\cite{HEND92}). Here, considering that the $b$-value
itself solely focuses on the one origin of self-similarity, and in
particular the process' increments infinite variance, we show
that, when employing natural time analysis, the $b$-value decrease
before large earthquakes reflects an increase of the fluctuations
of the order parameter of seismicity when approaching the critical
point (mainshock, see below). The whole precursory variation of
the order parameter fluctuations, however, is more complex since
it captures {\em both} origins. Temporal correlations between
earthquake magnitudes {\em also} play an important role in this
precursory variation, thus leading to more spectacular results
compared to the ones obtained when restricting ourselves to
traditional analysis of $b$-value alone.

For a time series comprising $N$ events, we define\cite{NAT01} the
natural time $\chi_k$ for the occurrence of the $k$-th event (of
energy $Q_k$) by $\chi_k = k/N$. We then study the evolution of
the pair ($\chi_k, Q_k$) or ($\chi_k, p_k$), where
$p_k=Q_{k}/\sum_{n=1}^{N}Q_{n}$ is the normalized energy released
during the $k$-th event.  The quantity $\Phi(\omega )$ is defined
by $\Phi(\omega)=\sum_{k=1}^N p_k \exp (i \omega \chi_k )$, where
$\omega$ stands for the natural angular frequency, and then
evaluate the real function $\Pi(\omega)=|\Phi(\omega)|^2$ in the
low frequency limit. By considering the Taylor expansion
$\Pi(\omega)=1-\kappa_1 \omega^2+ \kappa_2 \omega^4 + \dots$, we
find that the approach of a dynamical system to criticality (see
Chapter 8 of Ref.\cite{SPRINGER})  is identified by means of
$\kappa_1$, i.e.,
\begin{equation}\label{k1}
\kappa_1 = \langle\chi^2\rangle - \langle \chi \rangle^2 =
\sum_{k=1}^{N} p_k \chi^2_k-\left( \sum_{k=1}^{N} p_k
\chi_k\right)^2,
\end{equation}
which is the variance\cite{SPRINGER,NAT01,NAT05C} of natural time
weighted for $p_k$. When $Q_k$ are independent and identically
distributed positive random variables, we obtain the ``uniform''
(u) distribution of $p_k$, as it was defined in Ref.\cite{NAT03A}
(see also p.122 of Ref.\cite{SPRINGER}). In this case, {\em all}
$p_k$ vary around their mean value $1/N$ (cf. since
$\sum_{n=1}^{N} p_n=1$) and the quantity $\kappa_1$
results\cite{NAT03A} in $\kappa_u=1/12$ for large $N$.

In general, in a complex time series, in order to identify the two
origins of self-similarity by means of natural time analysis, we
focus on the expectation value $\mathcal{E}(\kappa_1)$ of the
variance $\kappa_1$ of natural time when sliding a natural time
window of length $l$ through a time series of $Q_k > 0$, $k=1, 2,
\ldots N$.

\begin{figure}
\includegraphics{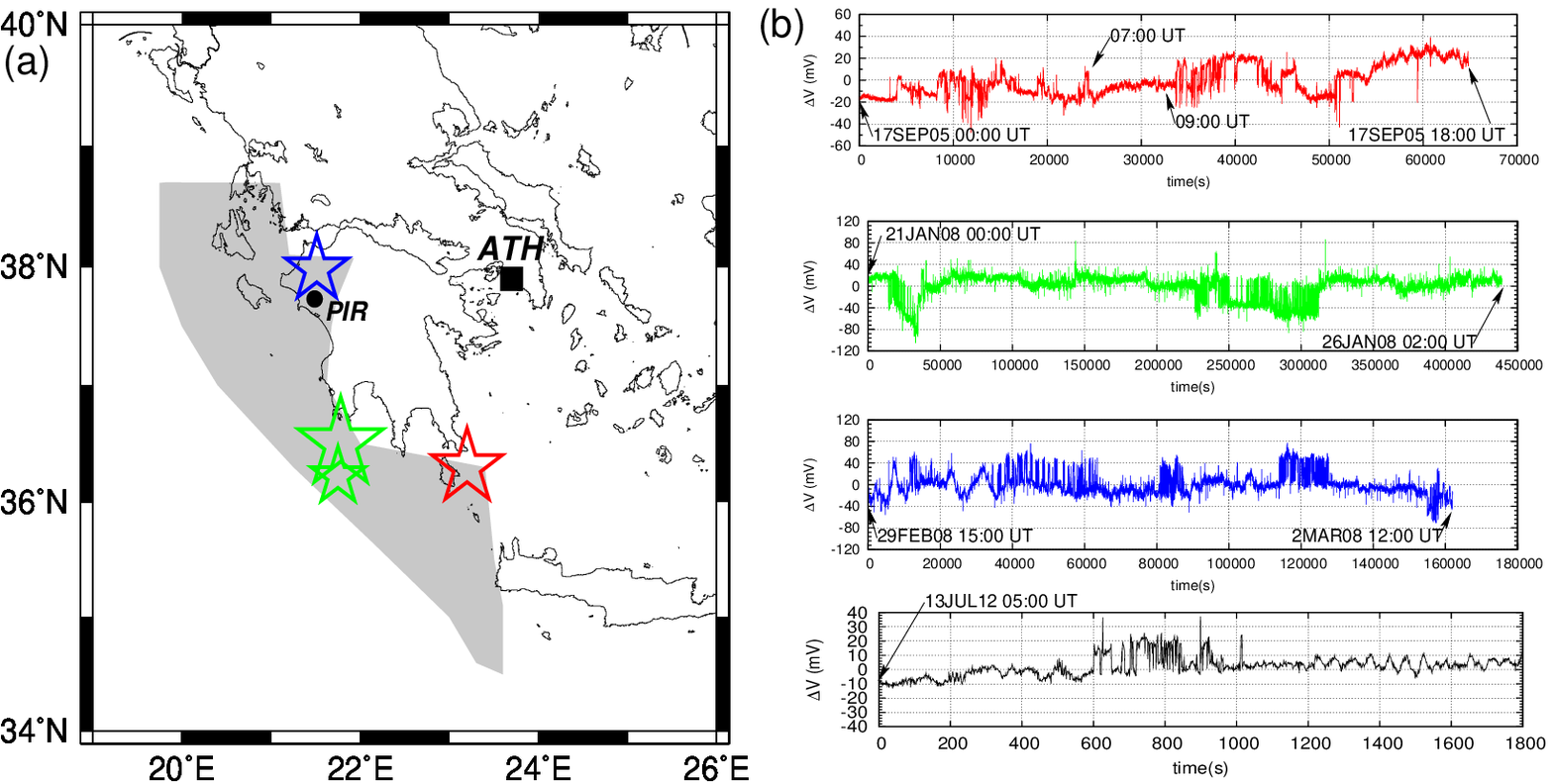}
\caption{(color online) (a) Major earthquakes in Greece on January
8, 2006 (red, magnitude $M_w=$6.7), February 14, 2008 (green,
$M_w=$6.9 and 6.4) and June 8, 2008 (blue, $M_w=$6.4) (b) Their
preceding SES activities recorded at Pirgos (PIR) measuring
station located in western Greece are shown (with the
corresponding color) in the upper three channels. Earthquakes with
SES activities at PIR are located in the shaded region of (a).
Furthermore, an SES activity recorded at a station in northern
Greece on July 13, 2012, is depicted in the lowest channel of (b).
}\label{f1}
\end{figure}

 If self-similarity exclusively results from the process' memory,
the $\mathcal{E}(\kappa_1)$ value should {\em change} to
$\kappa_u=1/12$ for the (randomly) shuffled data. This is the case
of the Seismic Electric Signals (SES) activities\cite{VAR93},
which are series of low-frequency ($\leq 1$Hz) electric signals
detected a few to several weeks (up to five months) before an
earthquake when the stress in the focal region reaches a {\em
critical} value (and hence long range correlations develop). For
example, the three upper channels in Fig.\ref{f1}(b) show three
SES activities that preceded major earthquakes in southern,
southwestern and western Greece, respectively, as depicted in the
map of Fig.\ref{f1}(a). For the sake of comparison, the lowest
channel shows an SES activity recorded in northern Greece (close
to Thessaloniki). In all these four cases, the analysis of their
original data lead to $\kappa_1\approx 0.07$ (see also below),
which turns to $\kappa_u=1/12$ upon shuffling the data.   On the
other hand, if the self-similarity results from process'
increments infinite variance {\em only}, $\mathcal{E}(\kappa_1)$
should be the same (but differing from $\kappa_u$) for the
original and the (randomly) shuffled data. Finally, when both
origins of self-similarity are present, the relative strength of
the contribution of the one origin compared to that of the other
can be quantified on the basis of Eqs.(12) and (13) of Ref.
\cite{NAT06B} (see also Ref.\cite{SPRINGER}).

In what remains, we focus on complex time series of seismicity.
Earthquakes exhibit scaling relations chief among which is the
aforementioned G-R law\cite{GUT54}. For reasons of convenience, we
write hereafter G-R law of Eq.(\ref{GRorig}) into the form $N(\geq
{\rm M})\propto 10^{-b {\rm M}}$. Considering that the seismic
energy $E$ released during an earthquake is related\cite{KAN78} to
the magnitude through $E \propto 10^{c {\rm M}}$, where $c$ is
around 1.5, the latter form turns to the distribution,
\begin{equation}\label{G-Rgamma}
    P(E) \propto E^{-\gamma}
\end{equation}
where $\gamma = 1 + b/1.5.$ Hence, $b \approx 1$ means that the
exponent $\gamma$ is around $\gamma$=1.6 to 1.7, see Table 2.1 of
Ref.\cite{SPRINGER}.

The complex correlations in time, space and magnitude of
earthquakes have been extensively
studied\cite{COR04,HOL06,EICH07,LIP09,LEN11}. The observed
earthquake scaling laws\cite{TUR97} seem to indicate the existence
of phenomena closely associated with the proximity of the system
to a {\em critical} point (e.g., see Ref. \cite{HOL06} and
references therein). In the frame of natural time analysis, it has
been suggested\cite{NAT05C} (see also pp.249-254 of
Ref.\cite{SPRINGER}) that the order parameter of seismicity is the
quantity $\kappa_1$. The $\kappa_1$ value itself may lead to the
determination of the occurrence time of the impending
mainshock\cite{SPRINGER,NAT01,NAT06A,NAT06B} when SES data are
available. In particular, when the $\kappa_1$ value resulting from
the natural time analysis of the seismicity subsequent to the SES
recording becomes approximately equal to 0.070, the mainshock
occurs within a time window of the order of one week. This has
been empirically observed in several
cases\cite{NAT01,NAT06A,NAT06B} (see also  Chapter 7 of Ref.
\cite{SPRINGER}) including the three major earthquakes of
Fig.\ref{f1}(a) that followed the SES activities depicted in
Fig.\ref{f1}(b). An example of the $\kappa_1$ dynamics after the
recording of the SES activity depicted in the third channel of
Fig.\ref{f1}(b) until the occurrence of the magnitude 6.4
mainshock on June 8, 2008 (blue star in Fig.\ref{f1}(a)) is given
in Ref.\cite{EPAPS}. In the lack of SES data, we have to solely
rely on the fluctuations of the order parameter of seismicity.
Along these lines, we investigated\cite{NEWEPL} the period before
and after a significant mainshock. Time-series for various lengths
of $W$ earthquakes that occurred before or after the mainshock
have been studied. The probability distribution function (pdf)
$P(\kappa_1)$ versus $\kappa_1$ was found to exhibit a bimodal
feature when approaching a mainshock. To quantify this feature, we
considered the {\em variability} of $\kappa_1$, which is just the
ratio \begin{equation} \beta \equiv \sigma( \kappa_1)/
\mu(\kappa_1), \end{equation} where $\sigma( \kappa_1)$ and
$\mu(\kappa_1)$ stand for the standard deviation and the mean
value of $\kappa_1$ for sliding window lengths $l$=6-40. The
bimodal feature reflects that, upon approaching the mainshock
(with the number $W$ of the earthquakes before mainshock
decreasing), the variability of $\kappa_1$ should increase. This
was subsequently confirmed because before the M9.0 devastating
Tohoku earthquake in Japan on March 11, 2011, the variability of
$\kappa_1$ exhibited\cite{JA} a dramatic increase.

In addition, we investigated\cite{EPL11B} the order parameter
fluctuations, but when considering a natural time window of a
fixed-length $W$ sliding through a seismic catalog (cf. in general
the results of  complexity measures when considering $W=$const
complement\cite{SPRINGER} those deduced when taking windows of
various lengths $W$). For earthquakes in California and Greece, we
found\cite{EPL11B} that when $W$ becomes compatible with the lead
time of the SES activities (i.e., of the order of a few months),
the fluctuations exhibit a global minimum before the strongest
mainshock that occurred during a 25- and 10-year period,
respectively.

\begin{figure}
\includegraphics{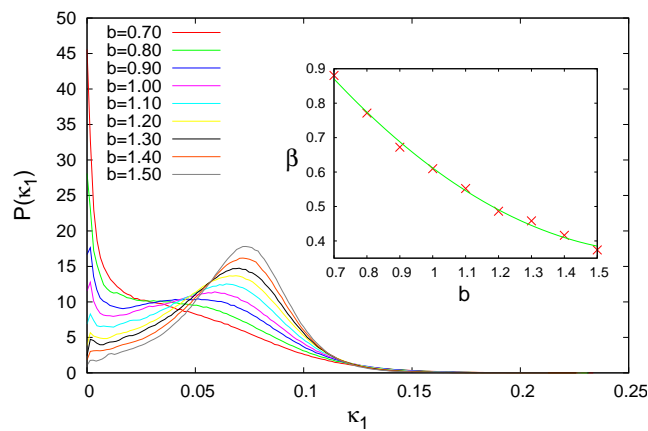}
\caption{(color online)The probability density function
$P(\kappa_1)$ versus $\kappa_1$ for several values of $b$ for
temporally uncorrelated events obeying Eq.(\ref{G-Rgamma}). The
inset depicts the variability $\beta$ as a function of $b$ (the
cross symbols refer to directly computed values, while the curve
has been drawn as a guide to the eye). }\label{f2}
\end{figure}

Let us now study the interrelation between the $b$-value and the
variability of $\kappa_1$. In particular, we investigate the
expected value of $\kappa_1$ when a natural time window length is
sliding through randomly shuffled power law distributed energy
bursts that obey Eq.(\ref{G-Rgamma}). In Fig.\ref{f2}, the pdf
$P(\kappa_1)$ versus $\kappa_1$ is plotted for several $b$ values,
an inspection of which reveals that: For high $b$-values, e.g.,
for $b$=1.5 and 1.4, the $P({\kappa_1})$ versus $\kappa_1$ curve
is almost unimodal maximizing at a value somewhat larger than
0.070, while for smaller $b$ a second mode emerges close to
$\kappa_1 \approx 0$ which reflects that the fluctuations of
$\kappa_1$ are larger. The computed values of the $\kappa_1$
variability as a function of the $b$ value are plotted in the
inset of Fig.\ref{f2}(b). The general feature of this curve is
more or less similar to that observed for example before Tohoku
earthquake\cite{JA}; quantitative agreement cannot be demanded,
however, because {\em temporal} correlations between the
earthquake magnitudes are also present which influence the
observed results. This is corroborated by the following results
obtained from the Olami-Feder-Christensen (OFC) earthquake
model\cite{OLA92}. We preferred to employ this model here, since
it has been studied in detail in hundreds of publications, but we
clarify that there exist more recent ones, e.g., see
Ref.\cite{DIET10} where the primary role of the fault system
geometry is emerged.

The OFC model runs as follows: we assign a continuous random
variable $z_{ij}\in(0,1)$ to each site of a square lattice, which
represents the local ``energy''. Starting with a random initial
configuration taken from a uniform distribution in the segment
(0,1), the value $z_{ij}$ of all sites is simultaneously increased
at a uniform loading rate until a site $i j$ reaches the threshold
value $z_{thres}$=1 (i.e., the loading $\Delta f$ is such that
$\left(z_{ij} \right)_{max}+\Delta f=1$). This site then topples
which means that $z_{ij}$ is reset to zero and an ``energy''
$\alpha z_{ij}$ is passed to every nearest neighbor, where the
coupling parameter $\alpha$ can take values from zero to 0.25 and
is the {\em only} parameter of the model, apart from  the edge
length $L$ of the square lattice. If this causes a neighbor to
exceed the threshold, the neighbor topples also, and the avalanche
continues until all $z_{k l} < 1$. Then the uniform loading
increase resumes. The number of topplings defines the size of an
avalanche or ``earthquake'' and (when it is larger than unity $k$
increases by one) is used as $Q_k$ in natural time analysis. Here,
we use the case of free boundary conditions\cite{HEL04} in which
$\alpha$ varies locally $\alpha_{ij} = \frac{1}{n_{ij}+K}$, where
$n_{ij}$ is the actual number of nearest neighbors of the site
$ij$ (for sites in the bulk $n_{ij}=4$, for sites at the edges
$n_{ij}=3$ and for the four sites at the corners $n_{ij}=2$) and
$K$ denotes\cite{HEL04} the elastic constant of the upper leaf
springs measured relatively to that of the other springs between
blocks in the Burridge-Knopoff model\cite{BUR67}. The OFC model is
obviously non-conservative for $K>0$ for which $\alpha_{ij}<0.25$
in the bulk (for more details on the OFC modelling see pp. 349-363
of Ref.\cite{SPRINGER} and references therein).

\begin{figure}
\includegraphics{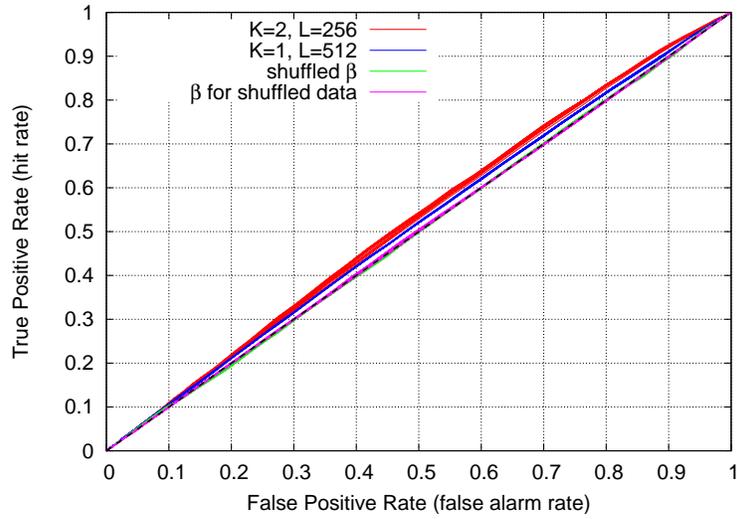}
\caption{(color online) The ROC diagram for the OFC earthquake
model discussed in the text: red ($L=256$ and $K=2$) and blue
($L=512$ and $K=1$) lines. In addition, two ROC diagrams are
depicted based on the results obtained for $L=512$ and $K=1$: The
green curves correspond to the case when the values of $\beta_k$
were randomly shuffled and the shuffled predictors were used,
while the magenta curves when the time-series of $Q_k$ was
randomly shuffled and then $\beta_k$ was estimated.}\label{f4}
\end{figure}

We first study the predictability of the OFC model on the basis of
the $\kappa_1$ variability. We consider the variability $\beta_k$
which is a function of the natural time index $k$, $k=1,2, \ldots,
N=2\times 10^6$ estimated by analyzing in natural time for each
$k$ the preceding $W=$100 avalanches.  The time increased
probability (TIP)\cite{KEI90A} (i.e., the time during which there
exists a high probability for the occurrence of a large avalanche
exceeding a given threshold) is turned on when $\beta_k>\beta_c$,
where $\beta_c$ is a given threshold in the prediction. If the
size $Q_k$ is greater than a target avalanche size threshold
$Q_c$, we have a successful prediction. For binary predictions,
the prediction of events becomes a classification task with two
types of errors: missing an event and giving a false alarm. We
therefore choose\cite{GAR09A} the receiver operating
characteristics (ROC) graph\cite{FAW06}  to depict the prediction
quality. This is a plot of the hit rate versus the false alarm
rate, as a function of the total rate of alarms, which here is
tuned by the threshold $\beta_c$. Only if in between the hit rate
exceeds the false alarm rate, the predictor is useful. Random
predictions generate equal hit and alarm rate, and hence they lead
to the diagonal in ROC plot. Thus, only when the points lie above
this diagonal the predictor is useful. As an example, the ROC
graphs for $L=512$ and $K=1$ or $L=256$ and $K=2$ are shown in
Fig. \ref{f4} (the rational for choosing these two cases stems
from the study of Ref.\cite{LIS01} in which it was shown that the
OFC model with free boundary conditions  exhibits  in these cases
-see their Fig.4- avalanche size distribution that agrees with the
G-R law). For every given threshold value $\beta_c$ and a target
threshold $Q_c$, we get a point in this plot, thus varying
$\beta_c$ we get a curve. The various curves in Fig. \ref{f4}
correspond to various values of $Q_c=168, \ldots, 1000$ increasing
from the bottom to the top. An inspection of this figure shows
that the points in each curve lie above the diagonal and the
excess is higher for larger values of $Q_c$. In order to
investigate the statistical validity of this result, we include in
the same graph the results where: (a) the values of $\beta_k$ were
randomly shuffled and the shuffled predictors were used (green
curves) and (b) the time-series of $Q_k$ was randomly shuffled and
then $\beta_k$ was estimated (magenta curves); in both cases, we
obtain curves which almost coincide with the diagonal. This
clearly demonstrates that the aforementioned excess of the results
related with the original $Q_k$ series from the diagonal comes
from the sequential order of avalanches and cannot be considered
as chancy.

\begin{figure}
\includegraphics{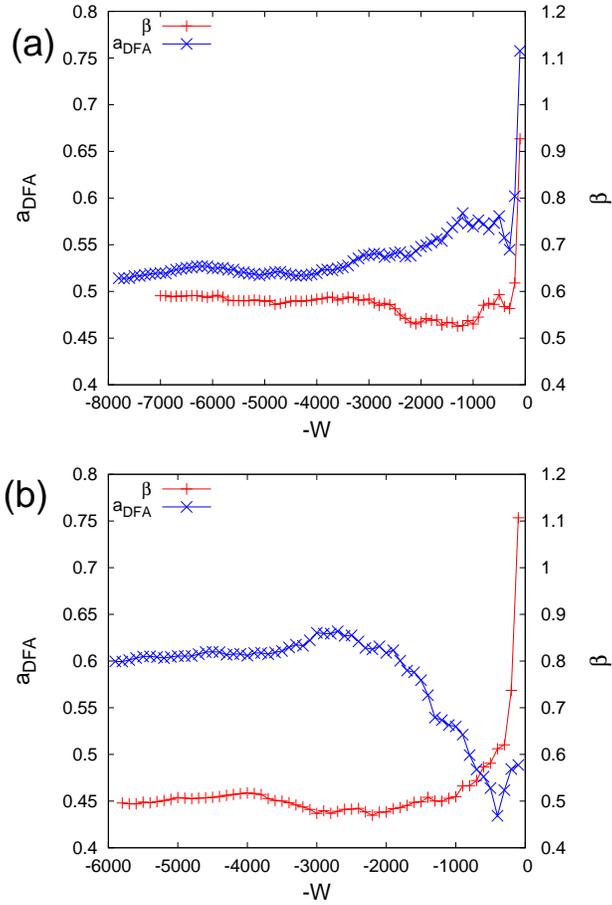}
\caption{(color online) The exponent $a_{DFA}$ (blue, left scale)
and the variability $\beta$ (red, right scale) versus the number
of the avalanches preceding a large avalanche, $Q_k=40,325$ for
(a) and $Q_k=31,145$ for (b), that corresponds to $W=0$ for the
OFC model ($K=1$, $L=512$).}\label{fg4}
\end{figure}

We now proceed to the investigation of the temporal correlations
between the magnitudes $m_k=log_{10}(Q_k)/1.5$ obtained from the
sizes $Q_k$ of the avalanches in the OFC model preceding a large
avalanche. The results can be visualized in two examples in
Fig.\ref{fg4}, where we plot in blue the exponent $a_{DFA}$ of the
detrended fluctuation analysis (DFA)\cite{PEN94} (along with the
variability $\beta$ plotted in red) versus the number $W$ of
avalanches before a large avalanche (negative $x$ semi-axis,
$x=-W$). Note that DFA has already been employed in
Ref.\cite{LIV07} for monitoring temporal correlations before
bifurcations.  In the upper example, Fig.\ref{fg4}(a), the value
of $a_{DFA}$ well before the large avalanche, being somewhat
larger than 0.5, exhibits small changes but strongly increases
upon approaching the large avalanche, i.e., at $W=100$ the value
of $a_{DFA}$ becomes $\approx0.75$ which shows intensified {\em
temporal} correlations. In the lower example, Fig.\ref{fg4}(b),
well before the large avalanche we have $a_{DFA}\approx0.6$
showing long range temporal correlations, which first turn to
anti-correlations upon approaching the large avalanche, e.g.,
$a_{DFA}\approx0.43$ at $W=400$, and finally become random, i.e,
$a_{DFA}\approx0.5$ at $W=100$, just before the ``mainshock''. In
both examples of Fig.\ref{fg4}, the variability $\beta$ rapidly
increases upon approaching a large avalanche  showing clear
precursory changes in the temporal correlations between
avalanches' magnitudes. A detailed statistical study of the OFC
model ($K=1$, $L=512$), for $W=100, 200, \ldots 1000$, showed that
among the 579 large ($Q_k > 30,000$) avalanches, only in 30\% of
the cases a rapid increase of $\beta$ upon approaching them is
observed. This is more or less consistent with empirical
observations since in Japan this precursory increase was observed
in 8 out of 25 earthquakes (all above M7 during 1 January 1994 to
11 March 2011 with depths smaller than 700 km)\cite{JA}.
Concerning the $\alpha$ values, when studying $W=100, 200, \ldots
1000$, among the 579 large avalanches studied, in 76\% of the
cases the $\alpha$ value was found to become smaller than 0.5 (as
seen in Fig.\ref{fg4}(b)).



\end{document}